\documentclass[11pt,english]{article}
\usepackage{newcent}
\usepackage[T1]{fontenc}
\usepackage[latin1]{inputenc}
\usepackage{geometry}
\usepackage{graphicx}
\usepackage{setspace}
\usepackage{natbib}
\providecommand{\tabularnewline}{\\}

\usepackage[pagebackref=true,colorlinks=true]{hyperref}
\usepackage{url}

\usepackage{babel}

\title{Efficient spike-sorting of multi-state neurons using inter-spike
intervals information}

\author{Matthieu Delescluse, Christophe Pouzat}

\begin{document}

\maketitle

\begin{abstract}
We demonstrate the efficacy of a new spike-sorting method based on
a Markov Chain Monte Carlo (MCMC) algorithm by applying it to real
data recorded from Purkinje cells (PCs) in young rat cerebellar slices.
This algorithm is unique in its capability to estimate and make use
of the firing statistics as well as the spike amplitude dynamics of
the recorded neurons. PCs exhibit multiple discharge states, giving
rise to multimodal interspike interval (ISI) histograms and to correlations
between successive ISIs. The amplitude of the spikes generated by
a PC in an {}``active'' state decreases, a feature typical of many
neurons from both vertebrates and invertebrates. These two features
constitute a major and recurrent problem for all the presently available
spike-sorting methods. We first show that a Hidden Markov Model with
3 log-Normal states provides a flexible and satisfying description
of the complex firing of single PCs. We then incorporate this model
into our previous MCMC based spike-sorting algorithm \cite[Pouzat et al,
2004, \emph{J. Neurophys.} \textbf{91}, 2910-2928]{PouzatEtAl_2004} and test this new
algorithm on multi-unit recordings of bursting PCs. We show that our
method successfully classifies the bursty spike trains fired by PCs
by using an independent single unit recording from a patch-clamp pipette. 
\end{abstract}
\pagebreak

\section{Introduction\label{sec:Introduction}}

Multi-site extracellular recordings are extensively used by laboratories
that aim at studying neuronal populations activity and a variety of
recently developed technologies enable the experimentalist to do so
in many preparations: cultures \cite{GrossEtAl_1993,GrossEtAl_1997}, slices
\cite{OkaEtAl_1999,EgertEtAl_2002}, \textit{in vivo}
\cite{DrakeEtAl_1988,NicolelisEtAl_1997,BakerEtAl_1999,CsicsavariEtAl_2003}.
But in order to be really informative and fully exploitable,
such recordings require the difficult spike-sorting problem to be
solved: the resolution of a mixture of activities into well separated
individual spike trains. This problem has an already long history
\cite{Lewicki_1998}, but has not yet received any fully satisfying solution
\cite{BrownEtAl_2004,Buzsaki_2004}. In particular, until recently
\cite{PouzatEtAl_2004,Pouzat_2005}, all of the available methods
made exclusively use of the information provided by the waveform of
individual spikes\footnote{Except \cite[Fee et al]{FeeEtAl_1996a} where the presence of a refractory period
was used in an \emph{ad hoc} way.}, ignoring that of their occurrence times. Many neurons have however
fairly reproducible firing features that can often be summarized by
their inter-spike interval (ISI) probability density. This temporal
information can greatly improve classification performance and allows
the investigator to take into account the dependence of the spike
amplitude upon the ISI, like for instance during a burst (where a
spike amplitude reduction is typically observed on an extracellular
electrode as well as on an intracellular one). 

In that context, we recently proposed a new Bayesian method based
on a Markov Chain Monte Carlo (MCMC) approach
\cite{PouzatEtAl_2004,Pouzat_2005}. This method is built on a data generation model that
includes both a description of non Poisson neuronal discharge statistics
and a description of spike waveform dynamics. In these papers, we
chose a \emph{single} log-Normal density to model individual ISIs
distributions. It is nevertheless clear that this is not the only
model that can be considered: the MCMC framework allows the experimentalist
to use the model that is best supported by the data from the neuronal
type he is studying. In particular, when one is dealing with neurons
exhibiting several states resulting in bursty discharges the unimodal
log-Normal density is not appropriate anymore, as observed for example
in thalamic relay cells \cite{McCormick_1998} and in cerebellar Purkinje
cells (PCs) \cite{LoewensteinEtAl_2005}. 

The primary goal of the present paper is to show how our spike-sorting
method, modified to take into account such multi-state neurons, performs
on real data that would make any other automatic method fail. We chose
a challenging data set recorded from cells firing bursts of spikes.
Such data could be obtained in young rat cerebellar slices by applying
a multi-site electrode along the PCs layer in the presence of the
group I metabotropic glutamate receptor (mGluR1) agonist (S)-3,5-dihydroxyphenylglycine
(DHPG). In these pharmacological conditions, PCs fire bursts of two
or more action potentials (APs) with dramatically decreasing amplitudes
(\textit{eg}. 50\%), in alternance with long periods of silence (on
the order of 1 minute) \cite{NetzebandEtAl_1997}.

\section{Methods\label{sec:Methods}}

\subsection{Experimental procedure\label{sub:Experimental-procedure}}

Slices preparation and loose cell-attached patch-clamp recording were
done as previously described \cite{PouzatHestrin_1997}. Sagittal
slices (180 $\mu$m thick) were taken from the vermis of cerebella
from rats aged 11-14 days. Single and multiple unit(s) recordings
were made from PCs in these slices visualized through a 63X objective
in an upright microscope equipped with Nomarski optics (Axioscope,
Carl Zeiss, Germany). These conditions allowed easy resolution of
the various layers and cell types within the cerebellar cortex. During
recording, the slices were maintained at room temperature (20-22 °C).
They were continuously perfused with BBS which contained (mM): 130
$NaCl$, 2.5 $KCl$, 2 $CaCl_{2}$, 1 $MgCl_{2}$, 1.3 $NaH_{2}PO_{4}$,
26 $NaHCO_{3}$ and 25 glucose. This solution was continuously bubbled
with a mixture of 95\% $O_{2}$ and 5\% $CO_{2}$, the solution pH
being thus kept at 7.4.

The mGluR1 agonist DHPG (40 $\mu$M) was applied dissolved in the
bathing solution. All the recordings analyzed in this paper were made
in these conditions. DHPG was purchased from Tocris Neuramin Ltd (Bristol,
UK). It was dissolved in distilled water at a concentration of 5 mM.
The DHPG stock solution was stored frozen, and the final concentration
was obtained by diluting the stock solution in the saline, just before
its use in the experiment. Note that spontaneous bursting also occurred
without DHPG at this age, but less systematically.

Single unit recordings were performed in loose cell-attached using
a glass micropipette filled with the following solution (mM): 145
$NaCl$, 2.5 $KCl$, 2 $CaCl_{2}$, 1 $MgCl_{2}$ , 10 HEPES acid.
Pipette resistance ranged from 2 to 4 M$\Omega$. The pipette was
positioned in loose cell-attached on the soma of a PC. It was connected
to a patch-clamp amplifier (Axoclamp 2B, Axon Instruments Inc., USA).
This amplifier was connected to one of the 8 channels of two 4-channel
differential AC amplifiers (AM systems, model 1700, Carlsborg, WA),
also used for multi-unit data recordings (see below). The signal was
band-pass filtered between 300 and 5000 Hz and amplified 1000 times.
Such single unit recordings were performed in two types of experiments.
First, they were made alone in order to gather several examples of
individual PCs spike trains during spontaneous activity. Second, they
were made together with multiple units recordings where they served
as a reference recording to which the spike-sorting output was compared.

Multi-unit recordings were performed using silicon probes (also called
``multi-site electrodes'' in the sequel) kindly provided by the
Center for Neural Communication Technology of the University of Michigan.
A schematic drawing of the tip (first 4 recording sites) of the probe
is shown in Fig.~\ref{fig:Sorting1}. The 16 recording sites are
linearly placed on the electrode 50 $\mu$m apart . This electrode
was positioned along the PCs layer. The spontaneous spiking activities
of these PCs could routinely be recorded on the first 8 sites of the
electrode, with an excellent signal-to-noise ratio (see Fig.~\ref{fig:Sorting1}B1,
B2, C1, C2 for some examples). The analysis detailed in the present
paper was made on the first 4 recording sites. A glass micropipette
was positioned in loose cell-attached on the soma of one of the PCs
whose spontaneous activity was recorded by the multi-site electrode.

The multi-site electrode was connected to a custom made impedance
matching preamplifier. The preamplifier was connected to the two 4-channel
differential AC amplifiers mentioned above. The signals were bandpass
filtered between 300 and 5000 Hz and amplified 2000 times. All data
were acquired at 15 kHz using a 16 bit A/D card (PD2MF-64-500/16H,
United Electronics Industries, Watertown, MA) and stored on disk for
subsequent analysis.

\subsection{Data analysis\label{sub:Data-analysis}}

\subsubsection{Events detection and representation\label{sub:Event-representation}}

\emph{Multi-unit data}. Data recorded on the first 4 recording sites
of the multi-site electrode were analyzed. A first set of large events
were detected as local maxima with a peak value exceeding a preset
high threshold (5 times the standard deviation (SD) of the whole trace),
and normalized (peak amplitude, at 1, temporal average, at 0) to give
a ``spike template''. Each trace was then filtered with this template
(by convolution with the template in reversed time order). Events
were detected on the filtered trace as local maxima whose peak value
exceeded a preset threshold (a multiple of the SD of the filtered
trace). After detection, each event was described by its occurrence
time and its peak amplitude measured on 4 recording sites. To simplify
calculations and reduce the complexity of our algorithm, the peak
amplitude(s) were ``noise whitened'' as described in
\cite{PouzatEtAl_2002} (see also \emph{SpikeOMaticTutorial
  1}\footnote{http://www.biomedicale.univ-paris5.fr/physcerv/C\_Pouzat/SOM.html}). A spike detected on a given recording site can be seen on its immediate
neighbouring sites ($50\mu m$ apart) with reduced amplitudes, but
never on further sites. This is consistent with an exponential decay
of the signal with decay constant $~30\mu m$ \cite{GrayEtAl_1995,SegevEtAl_2004}.

\emph{Single unit data}. For data recorded by the glass micropipette,
events were detected the same way. Each event was described by its
occurrence time and its peak amplitude. We normalized the peak amplitudes
of the spikes by the SD of 2000 noise ``peak'' amplitudes taken
in the same recording. When single unit data were recorded together
with multi-unit data, events detected on the micropipette trace were
described by their occurrence times only.

\subsubsection{Data generation model for statistical inference\label{sub:Data-generation-model}}

To perform spike-sorting our algorithm makes statistical inference
on the parameters of the data generation model described in this section.
This model is based on the following assumptions:

1. The sequence of spike times from a given neuron is a realization
of a Hidden Markov point process \cite{CamprouxEtAl_1996,GucluBolanowski_2004}.

2. The spike amplitudes generated by a neuron depend on the elapsed
time since the previous spike of this neuron.

3. The measured spike amplitudes are corrupted by a Gaussian white
noise which sums linearly with the spikes and is statistically independent
of them.

Assumptions~2 and~3 are identical to those made in
\cite{PouzatEtAl_2004,Pouzat_2005}. Assumption~3 requires a prior noise whitening
of the data \cite{PouzatEtAl_2002}. In assumption~1, the \emph{homogeneous
renewal point process} assumption that was made in
\cite{PouzatEtAl_2004,Pouzat_2005} is changed into the more complex one of a \emph{hidden
Markov point process} (see below and in Appendix sec.\ref{sub:More-Formal-HMM}).
Our data generation model can be divided in two parts that respectively
rely on assumptions~1 and~2 which are presented next.

\subsubsection*{Inter-spike interval density}

We resort to a Hidden Markov Model (HMM) with 3 states to account
for the empirical ISI density of the recorded cells. In this HMM context,
we can see a sequence of ISIs (a spike train) produced by a given
neuron as the observable output of a ``hidden'' sequence of ``states''
of this neuron (this denomination arises from the fact that the state
in which the neuron is, is not directly observable from the data).
The probability density from which each ISI is drawn depends on the
underlying state. In our particular implementation, the ISI density
of each state is a log-Normal density characterized by 2 parameters:
a scale parameter \textit{s} (in seconds) and a shape parameter \textit{$\sigma$}
(dimensionless). With this notation, the general formula for the probability
density function of the log-Normal distribution is:

\begin{equation}
f(isi)=\frac{1}{isi\cdot\sigma\cdot\sqrt{2\pi}}\cdot\exp\left(-\frac{1}{2}\cdot(\frac{\log(\frac{isi}{s})}{\sigma})^{2}\right)\label{eq:logNorm}\end{equation}

The log-Normal density is a relevant alternative to the exponential
density usually used to model spike trains. It is unimodal, exhibits
a refractory period, rises fast and decays slowly. 

After the generation of each event, a ``transition'' to any of
the three possible states is performed stochastically. In addition
to the 6 parameters for the 3 log-Normal densities mentioned above,
we have therefore to consider the transition matrix $\left(q_{ij}\right)$
between these states, which contains another 6 parameters. We thus
have 12 parameters to specify the ISI density for each neuron.

A scheme summarizing this model is shown in Fig.~\ref{fig:Model-ISI}A.
In a spike train, each event (ISI) is generated by one of the 3 possible
probability densities according to the state in which the neuron is:
if the neuron is in state~1 (resp.~2,~3) it generates a short (resp.
intermediate, long) ISI from the red (resp. blue, green) density.
The transition from a given state to any other, including itself,
is possible, as indicated by the different arrows between states.
In the sequel we will constantly refer to the same color code for
the states of single unit data: state~1 in red, state~2 in blue,
state~3 in green. They will be also called ``short'', ``intermediate''
and ``long'' states respectively. A more formal presentation of
the HMM is to be found in sec. \ref{sub:More-Formal-HMM} of the Appendix.

\subsubsection*{Spike amplitude dynamics}\label{sub:SpikeAmplitudeDynamics}

We use here the same spike amplitude dynamics as in
\cite{PouzatEtAl_2004,Pouzat_2005}. We consider events described by their occurrence
time and their peak amplitude measured on 1 (single site recordings)
or 4 recording sites (multi-site recordings). We model the dependence
of the amplitude on the ISI by an exponential relaxation \cite{FeeEtAl_1996b}:

\begin{equation}
A(isi)=P\cdot(1-\delta\cdot\exp(-\lambda\: isi))\label{eq:ampRelax}\end{equation}

where \emph{isi} is the ISI, $\lambda$ is the inverse of the relaxation
time constant (measured in 1/s), \textbf{$P$} is the vector of the
maximal amplitude of the event on each recording site (this is a 4-dimension,
resp. 1-dimension, vector for multi-site, resp. single site, recordings)
and $\delta$ is the dimensionless maximal modulation. This model
implies that the modulation of the amplitudes of an event is the same
on the different recording sites. This is an important feature of
the amplitude modulation observed experimentally \cite{GrayEtAl_1995}.
Added to the 12 parameters used for the ISIs, the number of parameters
\emph{per} neuron in our model amounts to 18 for multi-site recordings
and 15 for single site recordings.

\subsubsection{The Markov Chain Monte Carlo approach\label{sub:The-Markov-Chain}}

We have shown in our previous papers that the spike-sorting problem
with a data generation model similar to the one presented here can
be viewed as a one dimensional Potts spin-glass in a random magnetic
field \cite{PouzatEtAl_2004,Pouzat_2005}. This analogy allowed us
to tailor the Dynamic Monte Carlo algorithms developed by physicists
\cite{NewmanBarkema_1999,FrenkelSmit_2002} and the Markov
Chain Monte Carlo (MCMC) developed by statisticians
\cite{RobertCasella_1999,Liu_2001} for analogous problems to our particular needs. In essence the
statistical inference in our case relies on the construction of a
Markov chain (not to be mistaken for the HMM which is modeling the
ISI density) whose space $\mathcal{S}$ is the product of two spaces:
the space of the model parameters defined in our data generation model
( and presented in the previous section) and the space of spike train
configurations, where a configuration is defined by specifying a neuron
of origin (a ``label'') \emph{and} a neuron state for each spike.
The latter, neuron state, is the new model ingredient introduced in
the present paper (section \ref{par:Inter-spike-interval-density}).
Thus a state of our Markov chain in this space is determined by two
vectors: vector $\theta$ of model parameters (a $18\cdot K$, resp.$15\cdot K$,
dimensional vector for multi-site, resp. single site, recordings,
where $K$ is the number of neurons, see Methods, sec. \ref{sub:Data-generation-model})
and vector $C$ of the configuration, specifying a label and a neuron
state for each spike (a $2\cdot N$ dimensional vector, where $N$
is the number of detected spikes being analyzed). The construction
of this Markov chain is done in such a way that it samples our space
$\mathcal{S}$ from the posterior density of the model parameters
and configurations given the data $Y$, noted $\pi_{post}(\theta,C\mid Y)$:
at each step $t$ of the algorithm a new state $\left[\theta^{\left(t\right)},C^{\left(t\right)}\right]$
of the Markov chain (not to be mistaken for the neuron states of the
HMM used to model the ISI density) is generated from the state at
step $t-1$, $\left[\theta^{\left(t-1\right)},C^{\left(t-1\right)}\right]$,
according to the procedure described in Appendix, sec. \ref{sub:The-algorithm}.
The new components of the algorithm are the generation of the neuron
states of the HMM (when generating the new configuration) and the
generation of the transitions $q_{ij}$ between these neuron states
(which are components of the vector $\theta$ of parameters), using
a Dirichlet distribution \cite{RobertCasella_1999}. This way of generating
a new state from the previous one ensures that the Markov chain converges
to a unique stationary distribution given by $\pi_{post}(\theta,C\mid Y)$
\cite{PouzatEtAl_2004}.

As described in our previous model, the ``energy landscape'' explored
by our Markov chain exhibits some ``glassy'' features. It is therefore
necessary, in general, to use the Replica Exchange Method (REM, also
known as Parallel Tempering Method)
\cite{HukushimaNemoto_1996,Hansmann_1997,Iba_2001}  described in
\cite{PouzatEtAl_2004,Pouzat_2005}.
The method is not fully automatic yet and requires that the user chooses
the number of active neurons in the data by individually scanning
models with different numbers of neurons \cite{PouzatEtAl_2004}.

\subsubsection{Output analysis\label{sub:Bayesian-output-analysis}}

Once the simulated Markov chain has reached equilibrium, \emph{i.e.}
the chain is sampling from its stationary distribution which is our
desired posterior density, we can estimate the values of the parameters
and labels, as well as errors on these estimates. This is done by
averaging the value of a given parameter $\theta_{i}$ over the $N_{T}$
algorithm steps performed, after discarding the first $N_{D}$ steps
(\emph{burn-in}) \emph{}necessary to reach equilibrium:\begin{equation}
\overline{\theta_{i}}=\frac{1}{N_{T}-N_{D}}\sum_{{\scriptstyle t=N_{D}}}^{{\scriptstyle N_{T}}}\theta_{i}^{\left(t\right)}\label{eq:meanPara}\end{equation}

However the successive Markov chain states generated by our algorithm
are correlated, that is, the values of a given parameter $\theta_{i}$
at successive steps are not independent. Therefore the standard deviation
(SD) of this parameter must be corrected for this autocorrelation
\cite{Janke_2002,Sokal_1989}. As explained in detail by \cite[Janke]{Janke_2002}
the correction is made by multiplying the empirical variance $\sigma^{2}(\theta_{i})$
of this parameter by the \emph{integrated autocorrelation time} $\tau_{autoco}(\theta_{i})$:

\begin{equation}
\tau_{autoco}(\theta_{i})=\frac{1}{2}+\sum_{l=1}^{L}\frac{\rho(l;\theta)}{\rho(0;\theta)}\label{eq:autocoTime}\end{equation}

where $L$ is the lag at which $\rho$ starts oscillating around $0$
and $\rho$ is the autocorrelation function of $\theta_{i}$:\begin{equation}
\rho(l;\theta_{i})=\frac{{\textstyle 1}}{{\textstyle N_{T}-N_{D}-1}}\sum_{t=N_{D}}^{N_{T}}\left(\theta_{i}^{\left(t\right)}-\overline{\theta_{i}}\right)\left(\theta_{i}^{\left(t+l\right)}-\overline{\theta_{i}}\right)\label{eq:acf}\end{equation}

We then have: \begin{equation}
Var(\theta_{i})=2\cdot\tau_{autoco}\cdot\sigma^{2}(\theta_{i})\label{eq:varPara}\end{equation}

\subsubsection{Software availability}

Our codes are freely available (under the Gnu Public Licence) and
can be found together with tutorials on our web
site\footnote{\url{http://www.biomedicale.univ-paris5.fr/physcerv/C_Pouzat/SOM.html}}.
The data presented in this paper are also freely
available\footnote{\url{http://www.biomedicale.univ-paris5.fr/physcerv/C_Pouzat/Compendium.html}},
as well as a compendium which enables the interested reader to reproduce
the whole analysis detailed in the Results section.

\section{Results\label{sec:Results}}

We proceed in two steps. We first detail and justify the data generation
model we chose to account for PCs firing statistics. We performed
loose cell-attached recordings of PCs in cerebellar slices in the
presence of DHPG, and we show that a Hidden Markov Model (HMM) with
3 log-Normal states fits reasonnably well the ISI histograms of the
individual spike trains obtained. For such single neuron data, our
algorithm is based on the construction of a Markov Chain on the space
of the HMM parameters and single spike train configurations, where
a configuration is defined by specifying one of the 3 states for each
spike. A Monte Carlo (MC) simulation is then used to estimate the
posterior density of the HMM parameters and of single spike train
configurations.

The second step is the inclusion of this model of neuronal discharge
into our general spike-sorting algorithm before running it on multi-unit
recordings of bursting PCs. Besides our multi-site electrode positioned
along the PCs layer, a glass micropipette independently caught the
activity of one of these PCs in loose cell-attached. We can therefore
show that our spike-sorting method reliably isolates the activity
of this reference cell, although it is firing bursts of spikes with
decreasing amplitudes and exhibits a muti-modal ISI histogram.

\subsection{Single unit recordings of Purkinje cells in loose cell-attached\label{sub:Single-unit-recordings}}

We performed single unit recordings of spontaneously active PCs (\textit{n}
= 12) using a glass micropipette (2-4 M$\Omega$) in loose cell-attached
in the presence of bath-applied DHPG (40 $\mu$M). In these conditions,
PCs systematically fire bursts of variable lengths, in alternance
with long periods of silence (on the order of 1 minute). After detection
of the events, the inter-spike interval (ISI) histogram of each spike
train was plotted. All of them were multi-modal. They had a principal
mode corresponding to the most frequent ISI of the cell in normal
condition (typically 60 ms), as well as a mode at longer ISIs (hundreds
of ms) whose width was variable. The third mode at short ISIs (5 to
10 ms) corresponds to the ISIs which are found in these bursts of
two or more spikes. In such bursts the amplitudes of the spikes are
strongly reduced (see sec. \ref{sub:Spike-amplitudes-relax}). 

\begin{figure}
\begin{center}\includegraphics[scale=0.6,angle=-90]{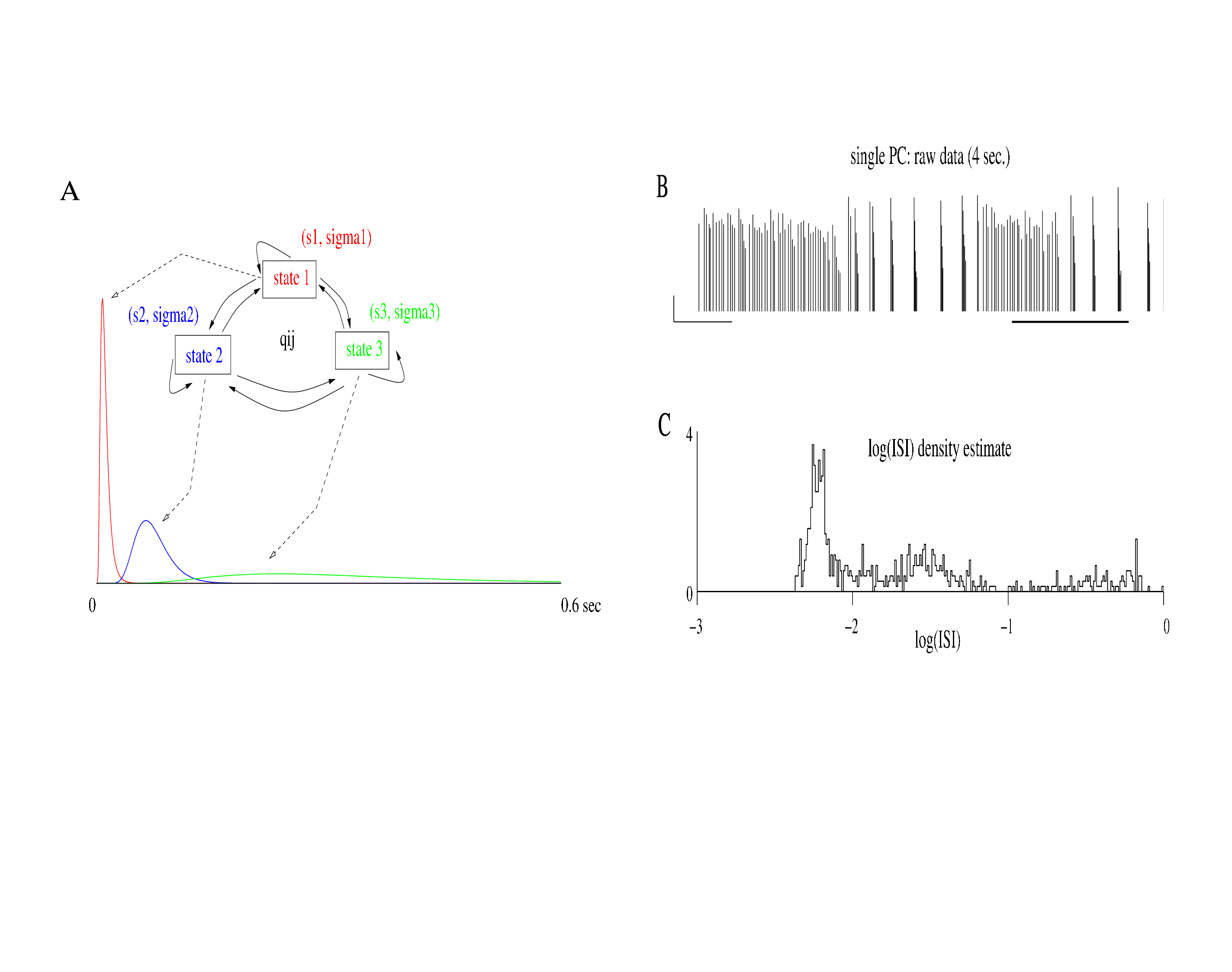}
\end{center}

\caption{\label{fig:Model-ISI}Model for the ISI density compared to a real
spike train.
A,~ISI density model: a Hidden Markov Model with three states. Every
state is a log-normal density with 2 parameters: a scale parameter\emph{~s}
(in seconds) and a shape parameter (dimensionless) $\sigma$: (0.01~sec.,~0.5)
state~1, (0.07~sec., 0.3) state~2, (0.3~sec.,~0.5) state~3.
B,~spontaneous activity of a single PC in presence of bath-applied
DHPG~(40~$\mu M$) in loose cell-attached. Normalized peak amplitudes
of the detected events are shown (duration: 4 seconds). The thick
horizontal bar on the right indicates the part of the train shown
in Fig.~\ref{fig:Model-ISI-Fit}B. Horizontal scale bar: 0.5~sec.
Vertical scale bar: 5 (in units of noise SD). C,~$\log_{10}\left(isi\right)$
histogram of the same spike train as in A (1~minute, 763~spikes).
Bin width:~0.01.}
\end{figure}

Fig.~\ref{fig:Model-ISI}B shows 4 seconds of a typical spike train
of a PC in DHPG. Note the presence of bursts of spikes of dramatically
decreasing amplitudes. The ISI histogram of this train (763 spikes,
1 minute of recording) is plotted in Fig.~\ref{fig:Model-ISI}C.
The multi-modal character of this histogram is unambiguous. This type
of activity (usually on the order of one minute) alternates with silent
periods with a similar duration of one minute.

\subsection{A 3-state Hidden Markov Model fits well empirical inter-spike interval
densities\label{sub:A-3-state-Hidden}}

\begin{figure}
\begin{center}\includegraphics[scale=0.5,angle=-90]{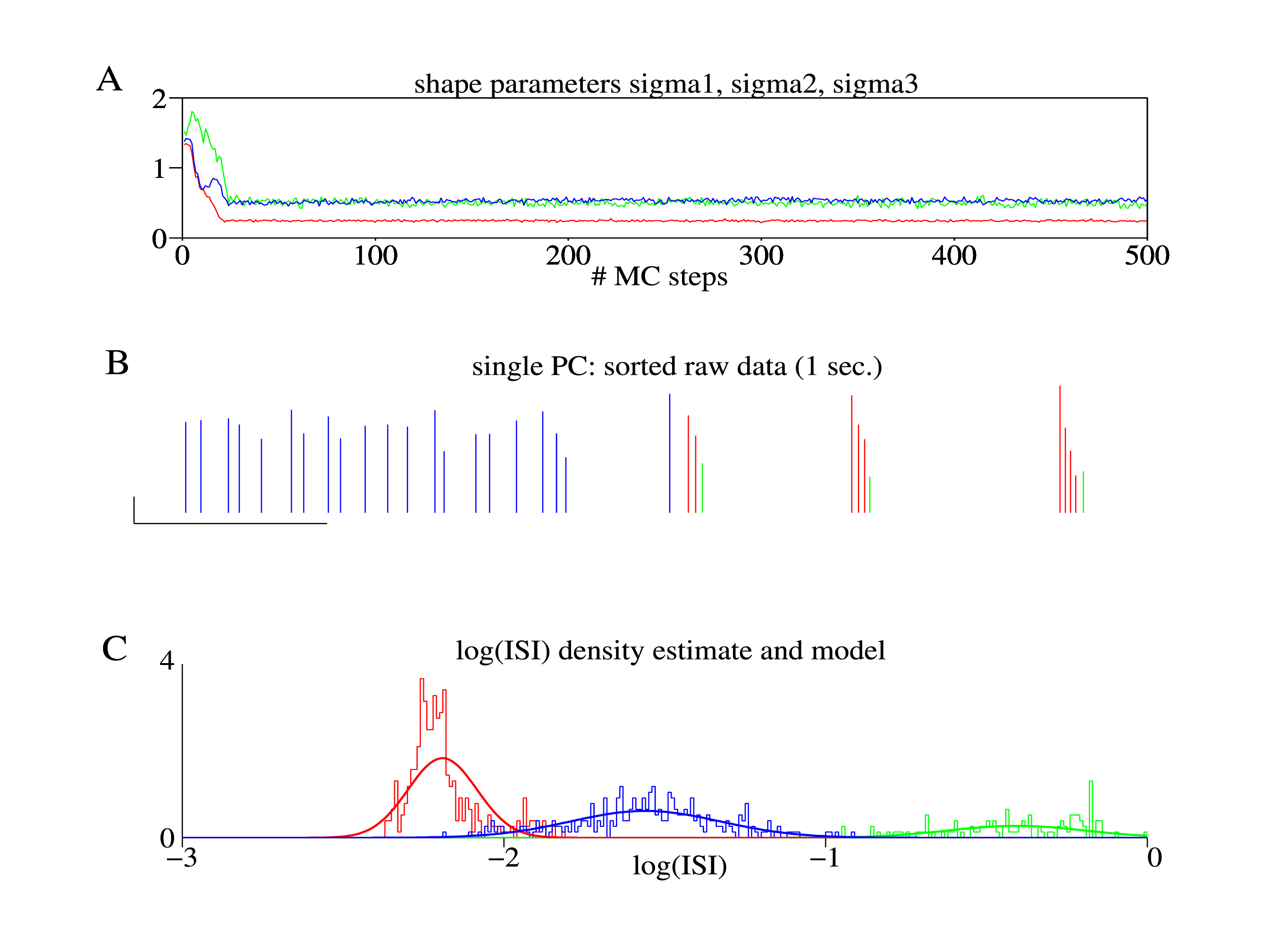}
\end{center}

\caption{\label{fig:Model-ISI-Fit}Sorting ISI modes in a cell engaged in
complex bursting behavior. 
\textsc{MCMC} output after a 1000-MC step run on the spike train shown
in Fig.~\textbf{\ref{fig:Model-ISI}.} A,~evolution of the shape
parameters \emph{$\sigma$$_{1}$,~$\sigma$$_{2}$,~$\sigma$$_{3}$}
during this 1000-step run. Only the first 500 MC steps are shown (same
color code as in Fig.~\ref{fig:Model-ISI}). B,~spike label analysis
of the episode indicated by the thick horizontal bar of Fig.~\ref{fig:Model-ISI}B.
Each spike is colored according to its most probable HMM state determined
by the algorithm (same color code as in Fig.~\ref{fig:Model-ISI}). Horizontal scale bar: 0.2~sec. Vertical
scale bar: 5 (in units of noise SD). C,~$\log_{10}\left(isi\right)$
histogram of each state for the most probable configuration (same
color code as in Fig.~\ref{fig:Model-ISI}). Note that each state corresponds to one mode of the histogram
of Fig.~\ref{fig:Model-ISI}C. The 3 model densities whose parameters
have been set at their average values computed on the last 200 iterations
are superimposed on the histogram.} 
\end{figure}

We used our MCMC algorithm without the REM (see Methods) to fit our
3-state HMM parameters, as well as the amplitude parameters (see sec.
\ref{sub:Spike-amplitudes-relax}, for the fit of amplitude parameters),
from this single unit spike train. In this section, where no spike-sorting
is performed, our algorithm only fits the model parameters of the
single cell recorded and attributes one of the three HMM states to
each spike in this single unit spike train. Fig.~\ref{fig:Model-ISI-Fit}A
shows the evolution of the dimensionless shape parameters $\sigma_{1}$~(red),
$\sigma_{2}$~(blue), \textit{$\sigma_{3}$}~(green) during a 1000-MC
steps run. Only the first 500 MC steps are displayed, but there was
absolutely no change in the evolution of these parameters between
steps 500 and 1000. All other parameters (scale parameters \textit{$\mathrm{s_{1}}$,
$\mathrm{s_{2}}$, $\mathrm{s_{3}}$} and transition parameters $\mathrm{q_{ij}}$)
had similar evolutions. Note that the algorithm reaches equilibrium
very fast (after about 20 MC steps). The average values of \textit{$\mathrm{s_{1}}$,
$\mathrm{s_{2}}$, $\mathrm{s_{3}}$} (autocorrelation corrected SDs
given in parenthesis, see Methods sec.~\ref{sub:Bayesian-output-analysis})
computed on the last 200 iterations were 6~(0.08)~ms, 28~(1)~ms,
392~(22)~ms respectively. These scale values are to be compared
to the location of the 3 ISI histogram's modes described in what follows.
The average values of $\sigma_{1}$, $\sigma_{2}$, \textit{$\sigma_{3}$}
(autocorrelation corrected SDs given in parenthesis, see Methods sec.~\ref{sub:Bayesian-output-analysis})
were 0.246~(0.01), 0.538~(0.026), 0.494~(0.041) respectively.

Fig.~\ref{fig:Model-ISI-Fit}B shows part of the spike train of Fig.~\ref{fig:Model-ISI}B.
At each step the algorithm attributes one of the 3 possible states
to each spike. The configuration (\textit{i.e}. the labeling of each
spike with a neuron state's number) shown in Fig.~\ref{fig:Model-ISI-Fit}B
is the most frequent one computed over the last 200 steps of the 1000-step
run displayed in A. We use the same color code as in Fig~\ref{fig:Model-ISI}A.
As expected, spikes in bursts are attributed by the algorithm to a
short state (red) label, except for the last spike of the burst which
is followed by a long ISI and is thus attributed to a long state (green)
label. Spikes occurring during regular spiking and separated by intermediate
ISIs are attributed the intermediate state (blue) label. 

In Fig.~\ref{fig:Model-ISI-Fit}C, the ISI histogram of this spike
train has been subdivided and colored according to the state of the
neuron: all ISIs generated when the neuron was in the short state
(resp. intermediate, long) are plotted in red (resp. blue, green),
as expected. Superimposed on this histogram are the 3 model ISI densities
whose parameters have been set to their average values computed on
the last 200 MC steps and given above. The reader sees that the initial
ISI histogram displayed in Fig.~\ref{fig:Model-ISI}C is reasonably
well fitted by this mixture of three log-normal densities (the most
striking deviation between the actual data and the fit being observed
for the short state). Moreover, besides its ability to describe a
multi-modal ISI histogram, the HMM can also account for the dependence
between successive ISIs through its transition matrix. In our case,
a long state is always followed by a short state, as shown more quantitatively
in sec. \ref{sub:Table-1:-Dependence} of the Appendix. 

\begin{figure}
\begin{center}\includegraphics[scale=0.6,angle=-90]{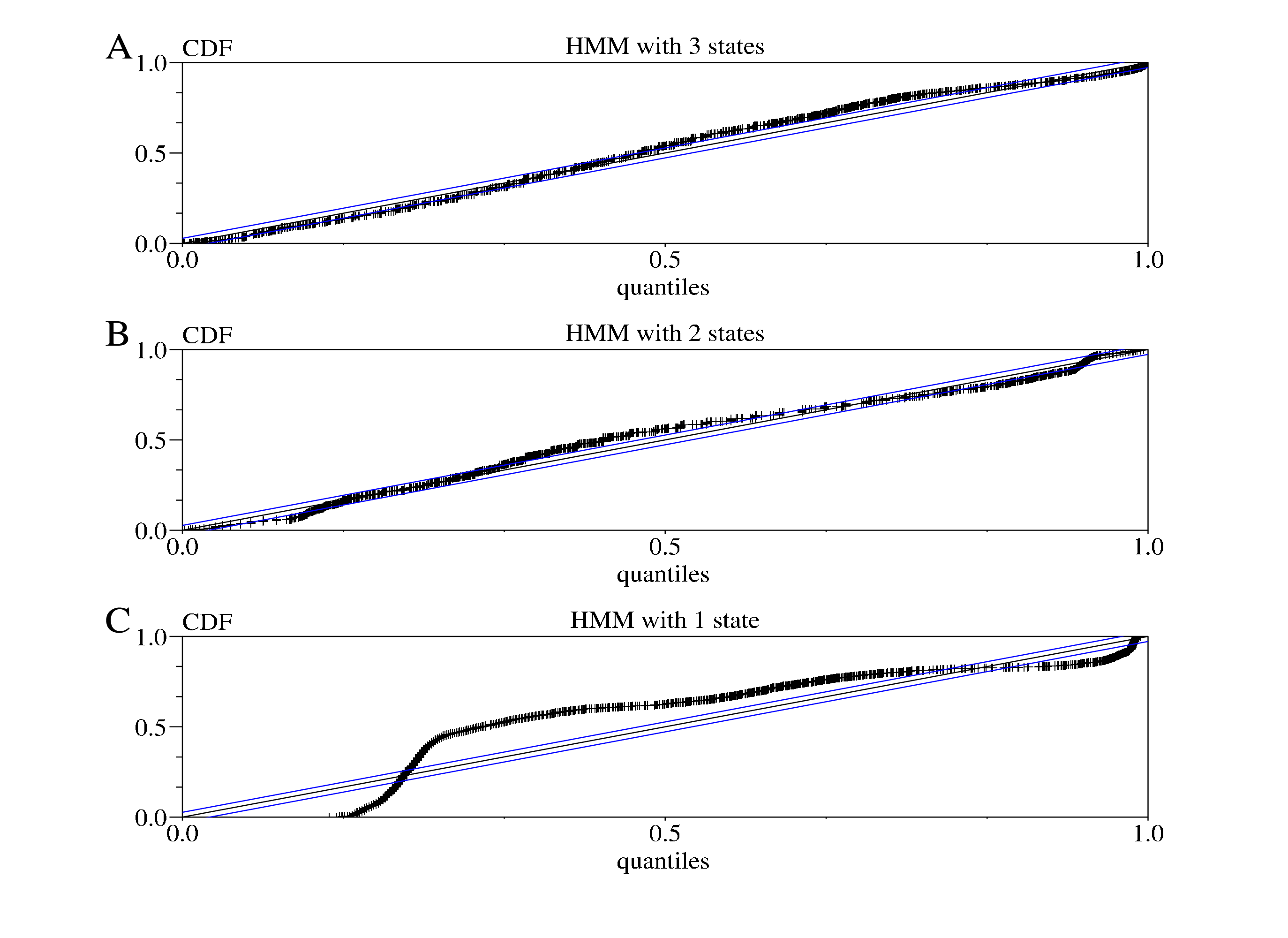}
\end{center}
\caption{\label{fig:K-S-test}Kolmogorov-Smirnov~(K-S) plots of the Hidden
Markov Models fits to the PC spike train shown in Fig. \ref{fig:Model-ISI}
and \ref{fig:Model-ISI-Fit}.
A,~HMM with 3~log-normal states. B,~HMM with 2~log-normal states.
C,~HMM with 1~log-normal state. In A, B and C, the solid 45-degree
line represents exact agreement between the model and the data. The
45-degree lines on both sides are the 95\% confidence bounds for exact
agreement between the model and experimental data based on the Kolmogorov-Smirnov
statistics. Although not perfect, a HMM with 3 states obviously fits
better the PC spike train than a HMM with 2 or 1 state(s). CDF stands
for Cumulative Distribution Function. For details about Kolmogorov-Smirnov
tests, see \cite{BrownEtAl_2001}.}
\end{figure}

These results show that the model we propose can satisfyingly, but
not perfectly, account for the discharge considered here. In Fig.~\ref{fig:K-S-test}
we provide the interested reader with a goodness-of-fit
test based on Kolmogorov-Smirnov plots. It shows that this spike train
clearly supports this 3 log-Normal state HMM when compared to models
with 2 and 1 state(s).

Between two periods of silence, a period of PC activity in DHPG always
evolves from a tonic firing at about 15~Hz (second mode of the ISI
histogram) to a bursty firing with 150~Hz-bursts (first mode of the
ISI histogram)\footnote{we are therefore approximating a non-stationary discharge dynamics
with a stationary one.} separated by intervals of several hundreds of ms (third mode of the
ISI histogram). This is well illustrated on Fig.~\ref{fig:Model-ISI}B
which represents a transition from this tonic to bursty firing. The
minute of activity shown in Fig.~\ref{fig:Model-ISI},~\ref{fig:Model-ISI-Fit}
and~\ref{fig:Amp-Dyn} depicts perfectly this evolution. However,
in the case where only the short bursts are recorded, only 2 modes
are prevailing (the short and the long ones). Such a case is obvious
in the multi-unit data analyzed in section \ref{sub:Multi-unit-data-sorted}.

\subsection{Spike amplitudes relax exponentially with respect to inter-spike
interval duration\label{sub:Spike-amplitudes-relax}}

\begin{figure}
\begin{center}\includegraphics[scale=0.6,angle=-90]{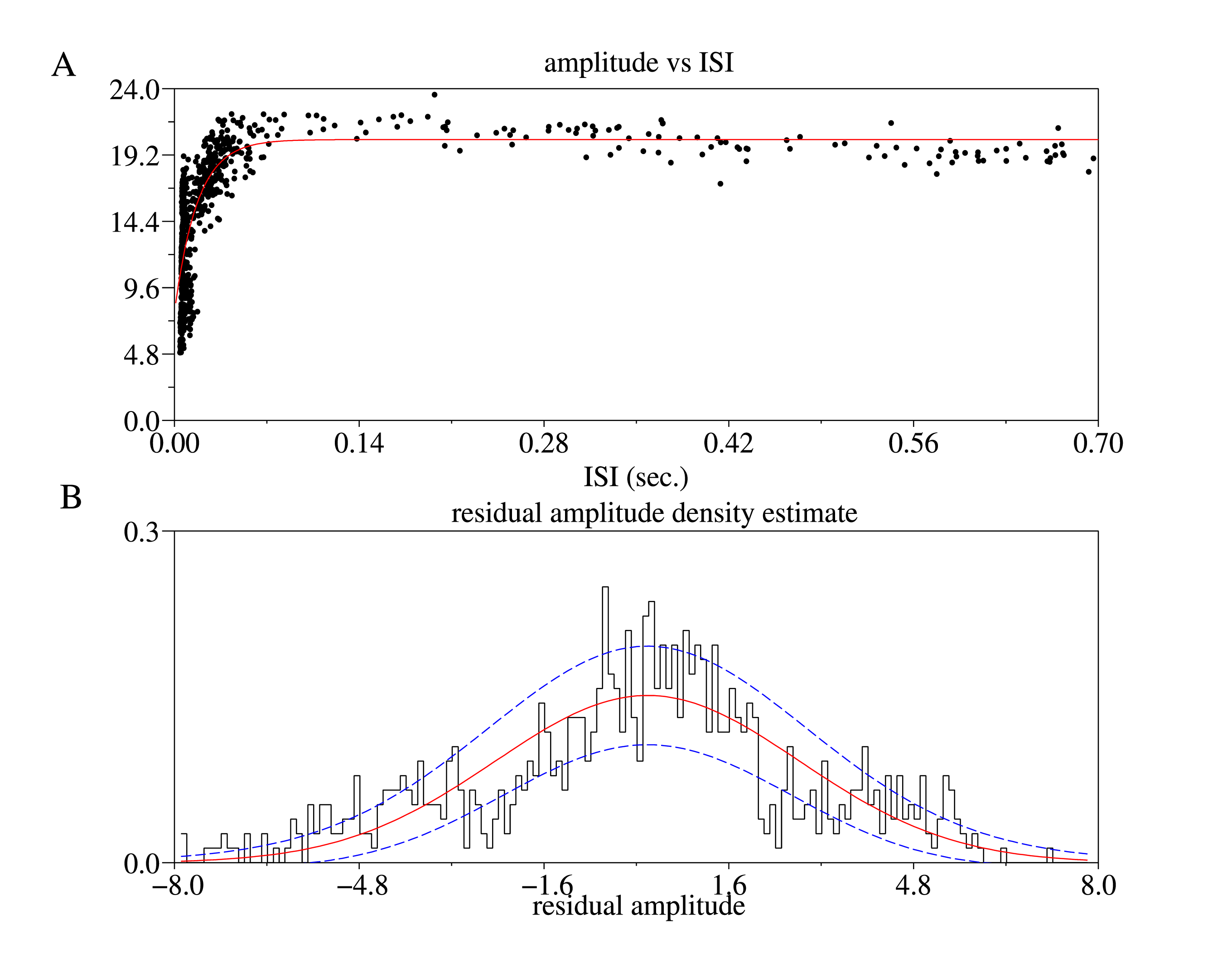}
\end{center}
\caption{\label{fig:Amp-Dyn}Model for the spike amplitude dynamics.
A,~Events normalized peak amplitudes plotted with respect to the
ISI preceding the event (dots). The fitted exponential relaxation
is superimposed (solid line). Ordinates in units of noise SD. B,~Histogram
of the residual amplitudes. The Gaussian fit is superimposed (solid
line). The two dashed lines are one SD away from the expected histogram.}
\end{figure}

The second part of our data generation model concerns the dependence
of spike amplitude upon the time elapsed since the last spike of the
same neuron. We also checked whether our PC spike trains supported
this hypothesis. We use here the same data set as in section \ref{sub:A-3-state-Hidden}.
In Fig.~\ref{fig:Amp-Dyn}A, the normalized peak amplitude of each
detected spike is plotted against the ISI that preceded it. Recall
that this single unit data were obtained with one recording site,
so that only one peak amplitude per spike is to be considered. The
exponential relaxation with parameter values determined by our algorithm
is superimposed on the data points. The reader is referred to section
\ref{sub:SpikeAmplitudeDynamics} where this exponential model is
presented. For this particular train, parameter values are (SDs given
in parenthesis): $P=20.3\,(0.03)$, $\delta=0.617\,(0.007)$, $1/\lambda=17\,(0.4)\, ms$.
$P$ is given in noise SD and $\delta$ is dimensionless. 

Several issues now must be addressed with respect to the peak amplitude
variance of the neurons we measured. First, the variability of spikes
amplitudes at short ISIs (around 5~ms) seems to be larger than those
at intermediate or long ISIs. This ``over-variability'' is mainly
a visual effect for a narrow range of short ISIs is significantly
more represented. This over-represented population necessarily samples
the Gaussian distribution more thoroughly. Second, a group of points
with an abscissa around 10~ms as well as points with abscissa greater
than 400~ms are clearly below the exponential fit, while points with
abscissa in 50-400~ms range are slightly above it. Third, these two
significant deviations compensate each other. This is shown by the
histogram of the residual amplitudes displayed in Fig.~\ref{fig:Amp-Dyn}B
to which a fitted Gaussian density with an SD equal to 2.64 is superimposed.
The SD of the residual amplitude histogram has been both added to
and subtracted from the fitting Gaussian curve (upper and lower dashed
line respectively). One sees that an exponential relaxation of amplitudes
looks like a good first approximation of the actual amplitude dynamics.
One sees as well that our third hypothesis is only an approximation,
for our residual here exhibit a larger variability (an SD of 2.64)
than the one expected from the measured background noise (SD of 1).
In particular, these data exhibit a slight but clear decrease of amplitude
with long ISIs, whereas our model keeps a fixed, maximal amplitude
for these ISIs.

\subsection{Multi-unit data sorted by our algorithm\label{sub:Multi-unit-data-sorted}}

\subsubsection{Multi-unit data and reference neuron\label{sub:Multi-unit-data-and}}

We performed multi-unit recordings of spontaneously active PCs in
the presence of DHPG using a multi-site electrode that we positioned
along the PC layer. A glass micropipette was placed in loose cell-attached
next to site~3 of the multi-site electrode in order to independently
monitor the activity of one of the PCs from the recorded population.
Such data were kept only if the glass micropipette unambiguously recorded
the activity of a single cell with an excellent signal-to-noise ratio.
This cell is called ``reference cell'' and the detected events
of these recordings serve as ``reference events'' to which the
output of our algorithm is compared. In what follows, we show the
performance of our spike-sorting algorithm on a representative example
of these recordings (58 seconds, 2739 events detected). Each detected
event is described by its time of occurrence and the 4 peak amplitudes
on the 4 recording sites after noise whitening.

\subsubsection{The reference neuron is reliably labeled as unit~1\label{sub:The-reference-neuron}}

The following results were obtained after a 1000-MC steps with the
REM and the following ``inverse temperatures'': $\beta=1,\,0.975,\,0.95,\,0.925,\,0.9,\,0.875,\,0.87$
followed by 1000 steps with $\beta=1$ only. This required about 33
minutes on a 3~GHz PC (Pentium IV) running Linux. Plots of energy
evolution and parameters evolution showed that all parameters had
reached their equilibrium value after roughly 500 MC steps (not shown).
We computed the average value of each model parameter using the last
200 MC steps. We also forced the soft classification produced by our
algorithm into the most likely classification using the last 200 MC
steps \cite{PouzatEtAl_2004}. 

\begin{figure}
\begin{center}\includegraphics[scale=0.5,angle=-90]{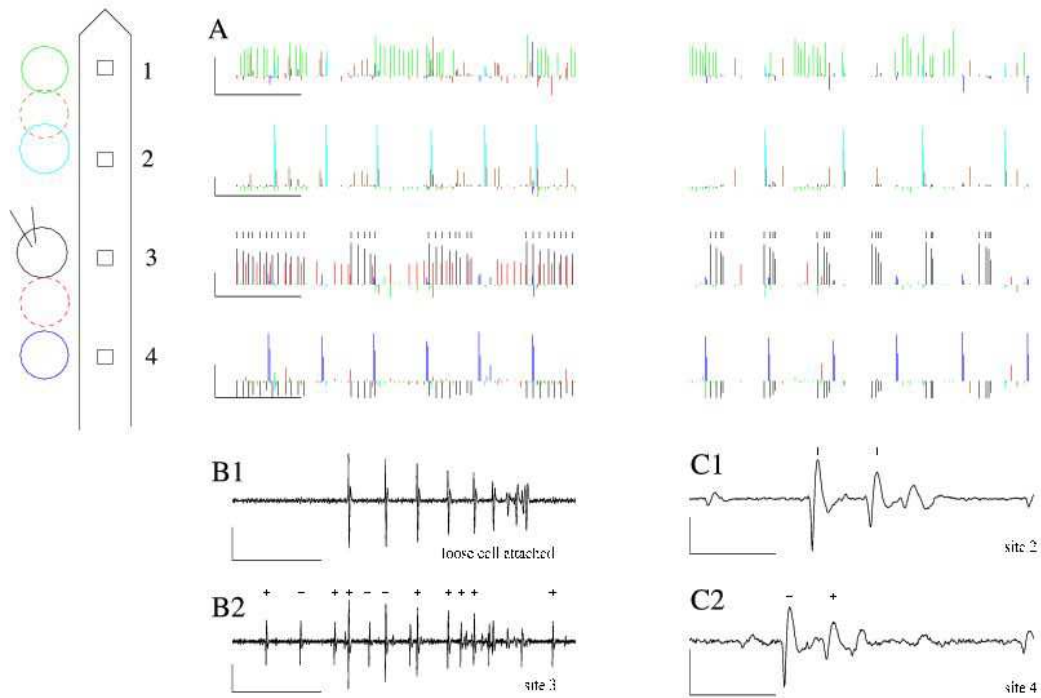}
\end{center}

\caption{\label{fig:Sorting1}Spike sorting on a mixture of several PC spike
trains simultaneously recorded (1).
A,~ spontaneous activity of PCs with bath-applied DHPG recorded on
the first 4 sites of the electrode. Two separate periods of 2~seconds
taken from the same minute of recording are shown (peak amplitudes
of the detected events after noise whitening). Each event is colored
according to the label determined by our algorithm. These labels are
the most probable ones over the last 200 steps of a 1000-step run.
An independent recording was performed next to site~3 by a patch-clamp
pipette in loose cell attached. A raster plot of this reference neuron's
activity is displayed in the upper part of the panel of site~3. A
scheme of the first 4 sites of the electrode as well as the recorded
PCs with the positioned extracellular glass pipette are drawn on the
left. Horizontal scale bar: 0.5~sec. Vertical scale bar: 10 (in units
of noise SD). B1: raw data recorded by the extracellular glass pipette
in loose cell attached, showing a typical burst. Horizontal scale
bar: 100~ms. Vertical scale bar: 0.5~mV. B2, corresponding raw data
recorded by site~3 of the multi-site electrode, showing the detected
burst. Crosses are drawn on top of the detected events (the same holds
for C1 and C2). Horizontal scale bar: 100~ms. Vertical scale bar:
0.25~mV. C1,~raw data recorded by site~2 of the multi-site electrode,
showing a typical triplet. Horizontal scale bar: 10~ms. Vertical
scale bar: 0.5~mV. C2,~raw data recorded by site~4 of the multi-site
electrode, showing a typical triplet (detected as doublet). Horizontal
scale bar: 10~ms. Vertical scale bar: 0.25~mV.}
\end{figure}

Fig.~\ref{fig:Sorting1}A shows two separate periods of 2 seconds
from these data (peak amplitudes of the detected events after noise
whitening, see Methods). Each row corresponds to one recording
site\footnote{in fact to a mixture of all of them (noise whitening has
  been performed), but the contribution of one site is still
  predominant.} of the Michigan probe (site~1 to site~4 from top to bottom), as
depicted on the left. Each event is colored according to its most
probable neuron of origin: neuron~1 in black, neuron~2 in deep blue,
neuron~3 in green, neuron~4 in light blue, neuron~5 in red, neuron~6
in brown. A raster plot of the reference events is displayed in the
upper part of the site~3 panel. From these plots, it is obvious that
the black unit (unit~1) reconstructed by our algorithm corresponds
to the reference cell (see details below). Note that the event amplitudes
of this cell are strongly reduced within the bursts so that they become
similar to those produced by unit~5 (red). This of course makes the
separation between these 2 units really difficult.

\begin{figure}
\begin{center}
\includegraphics[scale=0.5,angle=-90]{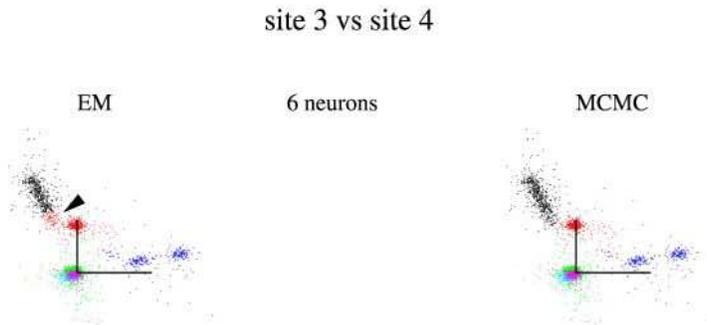}
\end{center}

\caption{\label{fig:Comp-EM-MCMC}Comparison between the sorting performed
by the Expectation Maximization (EM) algorithm and the MCMC algorithm.
Wilson plots (peak amplitudes of site~3 versus site~4) after sorting
based on a classic GMM(using the EM algorithm, left) and on our new
data generation model (using an MCMC algorithm, right) with 6 neurons.
The analysis performed with the MCMC algorithm for this number of
neurons is given in details in the article. The GMM fails to group
all the points of the reference neuron (black cluster), so that the
cluster is partially cut and attributed to the red one (arrow head):
the GMM attributes 84.5\% of the reference events to the reference
neuron (versus 98,1\% for the MCMC algorithm, see sec. \ref{sub:The-reference-neuron}).
This illustrates how inappropriate the GMM is with non-Gaussian, elongated
clusters. In such cases, implementing both the discharge statistics
of the neuron and its amplitude dynamics in the spike sorting method
very satisfyingly solves the problem.}
\end{figure}

Raw data from site~3 are displayed in Fig.~\ref{fig:Sorting1}B2
(crosses are drawn on top of the detected events), showing one of
the bursts of unit~1 as well as a ``background'' cell (for example,
the first three detected events on this panel come from this background
cell). As the spike amplitude decreases within the burst of unit~1,
these spikes end up being of the same size as this background cell.
The latter is labeled as unit~5 by our algorithm (red on panel A).
The corresponding raw data recorded by the independent micropipette
are displayed in Fig.~\ref{fig:Sorting1}B1. Note the huge signal
amplitude, as well as the decreasing spike amplitudes along the burst.
The activity recorded by the micropipette unambiguously comes from
a unique PC. These two panels B1 and B2 allow a direct comparison
of the signal received by site~3 of the microelectrode to the one
received by the pipette: the latter records the burst seen on panel
B2 only, and not the background cell. They also illustrate the fact
that not all events of the reference cell are detected on site~3:
the very last spikes of each burst fired by the reference cell are
much smaller and below our detection threshold. For that reason, among
the 766 reference events detected on the micropipette trace during
this minute of data, 641 are detected on the trace of site~3. Among
these 641 events, 629 are attributed to unit~1 by our algorithm (98,1\%).
Overall, 637 events are attributed to unit~1 so that 8 unit~1 events
are not reference events (false positives, 1.3\%). The 12 reference
events not labeled as unit~1 are labeled as unit~5 (red). For comparison,
a classical Gaussian mixture model (GMM) fitted with the Expectation-Maximization
(EM) algorithm (Pouzat et al., 2002), attributes only 542 reference
events to unit~1 (84.5\%), the 99 remaining ones being attributed
to unit~5. To illustrate this comparison, Fig. \ref{fig:Comp-EM-MCMC}
displays the Wilson plots, where the amplitude on site~4 is plotted
against its amplitude on site~3, after running the EM and the MCMC
algorithms separately: the Gaussian mixture model partially truncates
the elongated cluster of our reference neuron, whereas our elaborated
and more realistic model does not. This excellent performance of our
algorithm in such a difficult situation shows how powerful it is to
incorporate the temporal information into the spike-sorting procedure
through an appropriate model for the neuronal discharge statistics. 

Similar results were found in five other data sets of bursting PCs,
where the MCMC algorithm with the present HMM model outperformed the
EM algorithm. In three of these data sets, an extra pipette separately
recorded a reference cell: in these cases, our algorithm was able
to rebuild more than 96\% of the bursts of the reference cells, with
less than 3\% false positives.

\subsubsection{Unit~2 and unit~4 give rise to pairs of separated clusters on Wilson
plots\label{sub:Unit2-and-unit4}}

\begin{figure}
\begin{center}\includegraphics[scale=0.5,angle=-90]{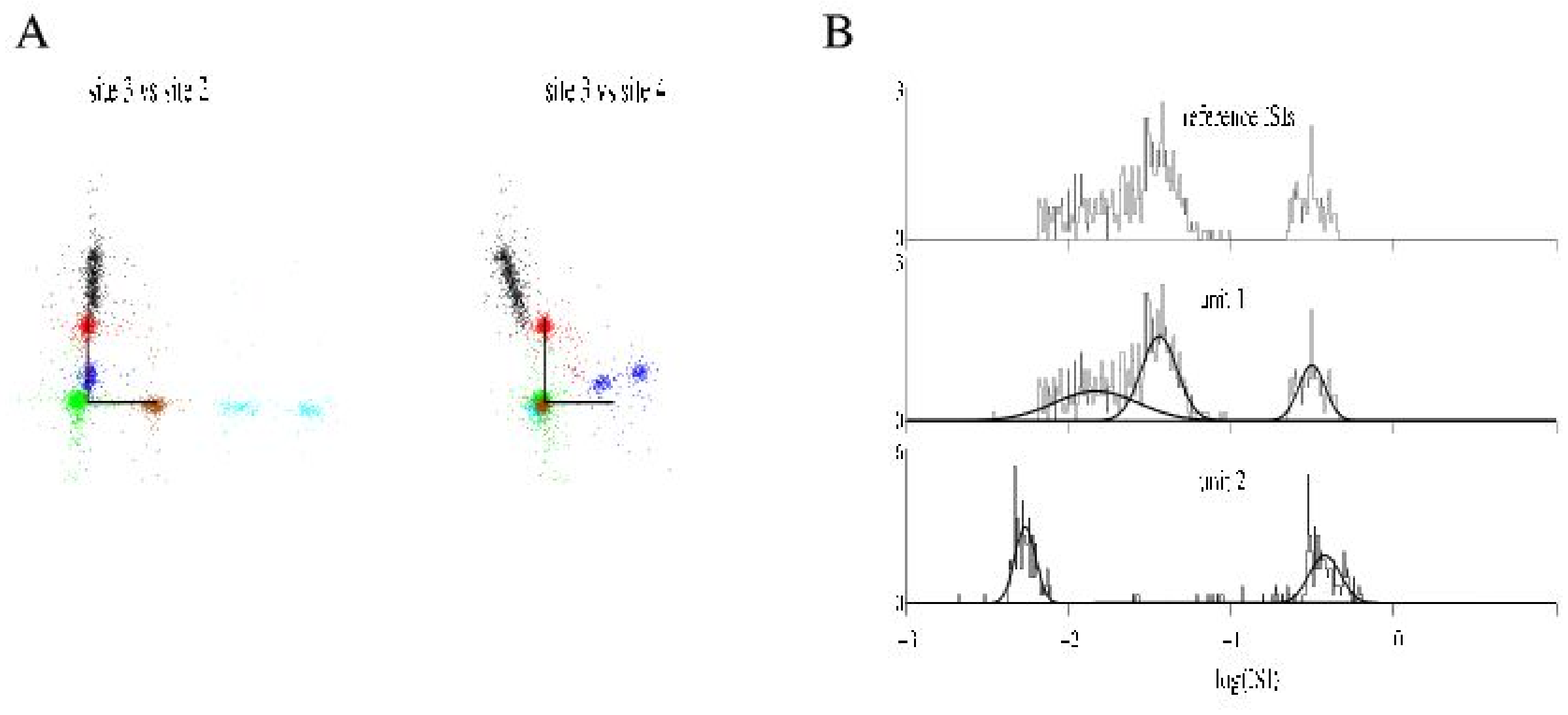}
\end{center}
\caption{\label{fig:Sorting2}Spike sorting on a mixture of several PC spike
trains simultaneously recorded (2).
A,~Wilson plots (peak amplitudes of site~3 against site~2 and site~4
respectively) showing the whole recorded sample (2739 events, 58 seconds).
Each event is colored according to the most probable label determined
by our algorithm (same colors as in Fig.~\ref{fig:Sorting1}A). The
reference unit is in black. Note the 2 pairs of separate clusters:
in light blue (site~2) and deep blue (site~4) on the lefthand and
righthand plots respectively. Note also the very elongated cluster
on both plots in black (site~3). On each plot the scale bars meet
at amplitude (0,0) and are of size 10 (in units of noise SD). B,~$\log_{10}(isi)$
histograms of units~1 (black cluster in A) and 2 (deep blue cluster
in A). Their 3 respective model densities whose parameters have been
set at their average values computed on the last 200 iterations are
superimposed on their respective histograms. The $\log_{10}(isi)$
histogram of the reference unit is also shown and to be compared to
unit~1 histogram.}
\end{figure}

Two other units deserve being examined. Unit~2 (deep blue, site~4)
and unit~4 (light blue, site~2) produce doublets of spikes of very
different amplitudes. Nevertheless, in both cases, these events are
recognized as coming from the same cell. The corresponding raw data
recorded on site~2 and site~4 are displayed in Fig.~\ref{fig:Sorting1}C1
and C2 respectively. These two panels show one typical burst of each
cell: in both cases, these bursts are in fact triplets of spikes.
In the case of site~4, the third spike of each burst remains below
detection threshold, so that only a doublet is detected (crosses on
top of the detected events). All these doublets are correctly identified
as coming from unit~2 (deep blue, Fig. \ref{fig:Sorting1}A). In
the case of site~2, the third spike of each burst is detected, but
is wrongly attributed to unit~6 (brown, Fig. \ref{fig:Sorting1}A),
instead of being attributed to unit~4, like the first two spikes
of the triplet (light blue, Fig. \ref{fig:Sorting1}A). This misclassification
is essentially due to the fact that our model of spike waveform dynamics
is not accurate enough for the data from this neuron, as discussed
in sec. \ref{sub:Why-misclassifications} and Fig.~\ref{fig:Amp-Dyn-Triplet}
in Appendix. This misclassification should moreover serve as a warning
against a blind use of our algorithm which would consist in taking
the output for granted without checking its relevance at all. The
plots displayed in Fig.~\ref{fig:Sorting1}A and Fig~\ref{fig:Sorting2}A,B
should be drawn after each run in order to assess the quality of the
sorting. In particular the ISI histograms of the sorted neurons must
show a clear refractory period and an overall shape that is similar
to ISI histograms of single cell recordings that can be obtained separately.

Fig.~\ref{fig:Sorting2}A shows Wilson plots of the data with the
same color code as in Fig.~\ref{fig:Sorting1}A. Only two plots out
of six are displayed. As in Fig~\ref{fig:Sorting1}A, unit~1 that
corresponds to the reference cell is in black. Note the elongation
of this cluster. Note also the 2 distinct, well separated clusters
of unit~4 (light blue) on the left hand plot.

\subsubsection{Empirical and modeled ISI densities\label{sub:Empirical-and-model}}

We finally display the ISI histograms of units~1 and~2, as well
as the one of the reference events detected on site~3 of the multi-site
electrode (Fig.~\ref{fig:Sorting2}B). The similarity between the
histogram of unit~1 and that of the reference cell is another illustration
of the 98\% performance of the algorithm on this unit. Note the 3
modes of this histogram. The 3 ISI model densities of unit~1 and
unit~2 with parameters set at their average values computed over
the last 200 MC steps are superimposed on their respective ISI histograms.
For unit~1, the 3 scale values are (autocorrelation corrected SDs
given in parenthesis) \{13~(1.6) ms, 35~(1.5) ms, 314~(5) ms\},
whereas the 3 shape values are \{0.395~(0.059), 0.262~(0.035), 0.194~(0.013)\}.
The scale values approximately locate the different modes of the histogram.
For unit~2, we have \{5~(0.06) ms, 92~(42) ms, 374~(12) ms) and
\{0.148~(0.009), 1.62~(0.479),0.224~ (0.028)\}. Only 20 events
of unit~2 (out of 301) are found to be in state~2. They correspond
to the few bins between the 2 modes of this histogram. As pointed
out in the next to last paragraph of section~\ref{sub:A-3-state-Hidden},
this unit only fires 150-Hz bursts during this minute of recording.
The more tonic firing that always occurs before was already over for
this unit by the time the recording started. This is why almost no
intermediate ISI is to be seen in this case. Each model density being
of course normalized by the proportion of events in each state for
a given unit, the curve corresponding to state~2 is almost null everywhere
and does not appear on the plot. This shows that, although the ISI
histogram of unit~2 is essentially bimodal, the behavior of the algorithm
is not altered at all. Our 3-state HMM can well accommodate any bi-
or unimodal ISI histogram. Like in section~\ref{sub:A-3-state-Hidden},
this shows how well the HMM accounts for the discharge statistics
of bursting cells that have tri- or bimodal ISI histograms.

\section{Discussion\label{sec:Discussion}}

We have shown here how the spike-sorting algorithm we recently proposed
\cite{PouzatEtAl_2004,Pouzat_2005}, modified for multi-state neurons,
performs on real, challenging data. In this data set, \emph{i.e.}
PCs in presence of DHPG, the recorded cells were firing bursts of
spikes whose amplitudes were strongly reduced, producing distinct,
well separated clusters in the Wilson plots (deep blue and light blue
clusters in Fig.~\ref{fig:Sorting2}), as well as very elongated
ones (black cluster in Fig.~\ref{fig:Sorting2}). To check the performance
of the algorithm, the activity of one of the recorded PCs was independently
and simultaneously monitored by a loose cell-attached glass micropipette
and served as reference spike train. We showed that our algorithm
did properly classify more than 98\% of this reference spike train,
despite the obvious decrease of spike amplitudes (Fig. \ref{fig:Sorting1}A,B1,B2
\& \ref{fig:Sorting2}A) and the tri-modal ISI histogram (Fig. \ref{fig:Sorting2}B).
We showed as well that it did associate the pairs of distinct clusters
mentioned above, obviously produced by a single neuron. In such situations,
existing methods require that the experimentalist \emph{a posteriori}
groups by himself the spikes that have been wrongly assigned to different
neurons due to their changing amplitudes. None of them can \emph{automatically}
give such an output on these data.

The excellent performance of our method relies on its ability to take
into account the information provided by the occurrence time of the
spikes, as well as their amplitude dynamics. To our knowledge, this
is the only method that makes use of these real spike trains properties.
It is moreover built on a proper probability model for data generation
which, in that case, implies that convergence proofs of the algorithm
do exist \cite{PouzatEtAl_2004,Pouzat_2005}. Our MCMC based approach
provides as well meaningful confidence intervals on the model parameters
and on the spike labels, a feature which should not be overlooked. 

We have also illustrated here the flexibility of the MCMC framework.
In our previous reports \cite{PouzatEtAl_2004,Pouzat_2005} we used
a simpler model for the discharge dynamics of the neurons: a single
log-normal density modeled the neurons ISI histograms. Here we first
showed that a multi-state HMM discharge model \cite{CamprouxEtAl_1996,
GucluBolanowski_2004} was well supported by our single unit
data (see sec. \ref{sub:A-3-state-Hidden}). We then included this
model in our spike-sorting algorithm and ran it on the multi-unit
data. The message is that once one knows how to write down an MCMC
algorithm for a ``reduced'' problem like generating the parameters
of the three states discharge model of a single neuron, it is straightforward
to incorporate it into the full spike-sorting algorithm. Therefore
if the experimentalist, based on single unit data (obtained for instance
with patch or sharp electrode recordings) thinks that another discharge
model would be better, the spike-sorting algorithm does not need to
be rewritten from scratch, only a sub-part of it needs to be modified.
We nevertheless think that the data generation model presented here
will turn out to be a good compromise between accuracy of the description
of real data and ease and speed of implementation. We did not seek
to relate each individual state of our HMM to any particular biophysical
event or set of events. This model has to be considered as a statistical,
descriptive tool that captures the key features of the observed neuronal
bursty firing. It is not limited to bursty firings though: in our
model, 3 states are available to describe the empirical ISI density,
but of course, one or two of these states can be unused for a unit
that has a uni- or bimodal ISI histogram (see unit~2 in section~\ref{sub:Empirical-and-model}).
Therefore, our model can account for uni- and bimodal ISI histograms
as well, of course better than a Poisson model would. In addition,
the number of states is not fixed at all and the experimentalist can
choose it himself as an input to the algorithm, \emph{a priori} and
for each neuron.

PCs are known to tonically fire action potentials as well as
bursts of spikes spontaneously in slices at 35 °C, even when
fast synaptic inputs are completely blocked
\cite{WomackKhodakhah_2002}. This spontaneous activity is preserved at room temperature
but bursts of spikes are less frequent. We facilitated the spontaneous
bursting behavior of PCs by adding DHPG to the bathing solution
\cite{NetzebandEtAl_1997}. This enables us to get multiunit
data in which most of the cells fire bursts of spikes of decreasing
amplitudes and helps demonstrate the ability of our spike-sorting
method to automatically isolate these bursts. 

For now the method is not fully automatic in the sense that it requires
the user to choose the number of neurons \emph{a priori} and give
it as an input to the algorithm. As discussed in \cite{PouzatEtAl_2004},
the general frame of the method provides a way to reliably compare
models with different numbers of neurons. This still ongoing work
will be reported in a near future.

\section*{Acknowledgments}

We thank Alain Marty and Ofer Mazor for comments and suggestions on
the manuscript. This work was supported in part by a grant from the
Ministère de la Recherche (ACI Neurosciences Intégratives et Computationnelles,
pré-projet, 2001-2003) and by a grant inter EPST (Bioinformatique).
Matthieu Delescluse was supported by a fellowship from the Ministère
de l'Education Nationale et de la Recherche.

Multichannel silicon probes were kindly provided by the University
of Michigan Center for Neural Communication Technology. The manuscript
was typed with LyX\footnote{\url{http://www.lyx.org}}. C codes were
developed in Emacs an debugged with
DDD\footnote{\url{http://www.gnu.org/software/ddd/}}.

\pagebreak

\pagebreak

\appendix

\section{Appendix\label{sec:Appendice}}

\subsection{Methods\label{sub:Methods}}

\subsubsection{A formal presentation of the HMM \label{sub:More-Formal-HMM}}

We extend here our previous model \cite{PouzatEtAl_2004,Pouzat_2005}
by introducing multiple \textsl{discharge states}, $D=d$, with $d\in\{1,\ldots,N_{ds}\}$,
for each neuron (in this paper we set the number of states $N_{ds}$
at 3). The goal is to be able to describe at the same time \textsl{multi-modal
ISI} densities (\textsl{i.e.}, densities with several peaks) and dependence
between successive \textsl{ISIs} similar to what goes on during a
burst, where (single) ``silent periods'' (long \textsl{ISIs})
are followed by many short \textsl{ISIs} (see sec. \ref{sub:Table-1:-Dependence}
for an example). Following \cite{CamprouxEtAl_1996} we assume that successive
\emph{ISI} are independent \emph{conditioned on the neuron's discharge
state}, $d$. After the emission of each spike (\textsl{i.e.}, the
generation of each \emph{ISI}) the neuron can change its \textsl{discharge
state} or keep the same. The inter discharge state dynamics is given
by a \textsl{Markov matrix}, $Q=(q_{ij})$. We moreover assume that
the \textsl{ISI} distribution of each state is \textsl{log-normal}.
In other words we assume that the neuron after its $m$th spike is
in state $d$ and that the \textsl{ISI} between this spike and the
next one is distributed as:

\begin{equation}
ISI\mid d\sim\textrm{log-normal}(s_{d},\sigma_{d})\label{eq:isi}\end{equation}

Eq.~\ref{eq:isi} should be read as:

\begin{equation}
\pi_{isi}(ISI=isi\mid S=s_{d},\Sigma=\sigma_{d})=\frac{1}{isi\cdot\sigma_{d}\cdot\sqrt{2\pi}}\cdot\exp\Big(-\frac{1}{2}\big(\frac{\log(isi)-log(s_{d})}{\sigma_{d}}\big)^{2}\Big)\label{eq:LogNor}\end{equation}

where $S$ is a \emph{scale} parameter (measured in seconds) and $\Sigma$
is a \emph{shape} parameter (dimensionless). These random variables
do depend on the value taken by the random variable $D$. After the
\textsl{ISI} has been generated, the neuron can ``move'' to any
of its $N_{ds}$ states according to a probabilistic dynamics described
by:

\begin{equation}
\textrm{P}(D^{(m+1)}=j\mid D^{(m)}=i)=q_{ij}\label{eq:InterStateDynamics}\end{equation}

You see therefore that if we work with a neuron with 3 discharge states
we have 12 independent \textsl{ISI parameters} to estimates: 2 pairs
$(s,\sigma)$ per state and $N_{ds}\cdot(N_{ds}-1)$ state transition
parameters (do not forget that matrix $(q_{ij})$ is \textsl{stochastic}
and therefore its rows sum to 1).

\subsubsection{Generating a new state in the Markov chain\label{sub:The-algorithm}}

In this section, we use $Y$ to designate the data. At step $t$,
state $\left[\theta^{\left(t\right)},C^{\left(t\right)}\right]$ is
drawn from state $\left[\theta^{\left(t-1\right)},C^{\left(t-1\right)}\right]$
by successively drawing each model parameter and each spike label
and spike state according to the procedures described below. To simplify
notations we omit the step index $(t)$ of the generated state. We
note $C_{-i}$ the configuration specifying the labels and neuron
states for all the spikes except spike number $i$. Similarly we note
$\theta_{-a}$ the vector of all model parameters except parameter
$a$. Each parameter $a$ has a uniform prior on a defined segment
$\left[a_{min},a_{max}\right]$ relevant for this parameter
\cite{PouzatEtAl_2004}. A step of our algorithm is performed once all spike
labels and states, as well as all model parameters have drawn as specified
below. This defines the new state in the Markov chain.

\subsubsection*{Labels and neuron states}

For each spike $i$ of the spike train, a label $l_{i}\in\left\{ 1,...,K\right\} $
and a neuron state $m_{i}\in\left\{ 1,2,3\right\} $ are drawn from
their posterior conditional density:

\begin{equation}
l_{i},m_{i}\mid Y,\theta,C_{-i}\sim\frac{\pi_{post}(l_{i},m_{i}\mid Y,\theta,C_{-i})}{\sum_{l_{j},m_{j}}\pi_{post}(l_{j},m_{j}\mid Y,\theta,C_{-i})}\label{eq:labelsDraw}\end{equation}

\subsubsection*{Amplitude parameters}

For each neuron $q$, the amplitude parameters ($P_{q,1},P_{q,2},P_{q,3},P_{q,4},\delta_{q},\lambda_{q}$)
are drawn with a Metropolis-Hastings step, using piecewise linear
approximations of their respective posterior conditional densities
as proposals \cite{PouzatEtAl_2004}. Let us take the case of $\lambda_{q}$,
for example, to illustrate the procedure. Let $\pi_{post}\left(\lambda_{q}\mid Y,\theta_{-\lambda_{q}},C\right)$
be its posterior conditional density and $\tilde{\pi}_{approx}\left(\lambda_{q}\mid Y,\theta_{-\lambda_{q}},C\right)$
be its piecewise linear approximation. Let $\lambda$ be the current
value of $\lambda_{q}$. 

First, $\tilde{\lambda}$ is drawn from the proposal density:

\begin{equation}
\tilde{\lambda}\mid Y,\theta_{-\lambda_{q}},C\sim\tilde{\pi}_{approx}\left(\lambda_{q}\mid Y,\theta_{-\lambda_{q}},C\right)\label{eq:ampDraw}\end{equation}

Then, this value $\tilde{\lambda}$ is accepted with probability $A$
equal to:

\begin{equation}
A=\min\left(1,\frac{\pi_{post}\left(\tilde{\lambda}\mid Y,\theta_{-\lambda_{q}},C\right)\cdot\tilde{\pi}_{approx}\left(\lambda\mid Y,\theta_{-\lambda_{q}},C\right)}{\pi_{post}\left(\lambda\mid Y,\theta_{-\lambda_{q}},C\right)\cdot\tilde{\pi}_{approx}\left(\tilde{\lambda}\mid Y,\theta_{-\lambda_{q}},C\right)}\right)\label{eq:acceptAmpDraw}\end{equation}

If $\tilde{\lambda}$ is accepted, then $\lambda\rightarrow\tilde{\lambda}$.

Else $\lambda\rightarrow\lambda$

\subsubsection*{Log-Normal parameters}

\emph{scale parameters}

For each neuron $q$ and each neuron state $r$ ($r\in\left\{ 1,2,3\right\} $)
of neuron $q$, the scale parameter of the log-Normal density is noted
$s_{q}^{r}$. 

First $u$ is drawn from \cite{PouzatEtAl_2004}:

\begin{equation}
u\mid Y,\theta_{-s_{q}^{r}},C\sim Norm\left(\overline{\log i_{q}},\frac{\left(\sigma_{q}^{r}\right)^{2}}{n_{q}}\right)\label{eq:scaleDraw}\end{equation}

where $n_{q}$ is the number of ISI of neuron $q$, $\sigma_{q}^{r}$
is the shape parameter of neuron $q$ in neuron state $r$, and $\overline{\log i_{q}}=\frac{1}{n_{q}}\sum_{j=1}^{n_{q}}\log i_{q,j}$,
$i_{q,j}$ being the ISI index $j$ of neuron $q$. 

Then, if $s=\exp\left(u\right)\in\left[s_{min},s_{max}\right]$, we
set $s_{q}^{r}=s$.

Else we draw another $u$. 

\emph{shape parameters}

For each neuron $q$ and each neuron state $r$ ($r\in\left\{ 1,2,3\right\} $)
of neuron $q$, the shape parameter of the log-Normal density is noted
$\sigma_{q}^{r}$. 

First $u$ is drawn from \cite{PouzatEtAl_2004}:

\begin{equation}
u\mid Y,\theta_{-\sigma_{q}^{r}},C\sim Gamma\left(\frac{n_{q}}{2}-1,\frac{n_{q}}{2}\left(\overline{\log i_{q}}-\log s_{q}^{r}\right)^{2}\right)\label{eq:shapeDraw}\end{equation}

with the same notations as for the scale parameters. 

Then, if $\sigma_{min}\leq\sqrt{1/u}\leq\sigma_{max}$, we set $\sigma_{q}^{r}=\sqrt{1/u}$.

Else we draw another $u$.

\subsubsection*{Transition parameters}

For each neuron $q$, the transition parameters between the 3 HMM
neuron states form a 9 by 9 matrix whose 3 rows are drawn successively.

Let $m=(m_{1},...,m_{N})$ be the spike train configuration of a neuron
$q$, where $m_{k}$ is the neuron state of spike $k$ of this neuron.
Let $\mathrm{n_{ij}}$ be the number of spikes of this neuron which
are in state $m_{j}$ following a spike of this neuron in state $m_{i}$.
The row number \emph{i} of the transition matrix is then drawn from
the Dirichlet distribution $\mathrm{\mathcal{D}_{3}}(1+n_{i1},1+n_{i2},1+n_{i3})$
\cite{RobertCasella_1999}.

\subsubsection{Implementation details\label{sub:Implementation-details}}

Codes were written in C. We used the free softwares
Scilab\footnote{\url{http://scilabsoft.inria.fr}} and
R\footnote{\url{http://www.r-project.org}} to generate output plots as
well as the graphical user interface.
The GNU Scientific
Library\footnote{\url{http://sources.redhat.com/gsl}} (GSL) was used
for vector and matrix manipulation  routines and (pseudo-)random
number generators. More specifically, the GSL implementation of the
Mersenne Twister of Matsumoto and Nishimura
\cite{MatsumotoNishimura_1998} was used to generate random variates. Codes were compiled with
the \emph{intel} icc
compiler\footnote{\url{http://www.intel.com/software/products/compilers/}}
and run on a PC (Pentium IV 3 GHz) running Linux.

\subsection{Supplementary analysis\label{sub:Supplementary-data-and}}

\subsubsection{Dependence between successive ISIs in the single unit spike train.\label{sub:Table-1:-Dependence}}

The transition matrix $\left(\mathrm{q_{ij}}\right)$ of the most
probable configuration (\emph{i.e} the attribution of a neuron state
to each spike) of the single unit spike strain shown in Fig. \ref{fig:Model-ISI}
and \ref{fig:Model-ISI-Fit} is given below. The lowest and largest
values taken by each transition element over the last 200 MC-steps
are given in square brackets. The neuron state numbers (\emph{i.e}
here, the row and column numbers) are those of Fig. \ref{fig:Model-ISI}
and \ref{fig:Model-ISI-Fit}, that is: states 1, 2 and 3 for the short,
intermediate and long ISIs respectively. The dependence between ISIs
is obvious: a long ISI is always followed by a short ISI ($\mathrm{\mathrm{q_{31}=1}}$),
a short ISI is either followed by another short ISI (within a burst),
or by a long ISI almost exclusively ($\mathrm{q_{11}=0.69}$ and $\mathrm{q_{13}=0.3}$).
This is in agreement with the existence of bursts separated by longer
intervals. This may be related to the refractory period after high
frequency discharge in burst. If successive ISIs were independent,
rows would be identical, each column being equal to the proportion
of the state.

\begin{center}\begin{tabular}{|c|c|c|c|}
\hline 
&
state 1&
state 2&
state 3\tabularnewline
\hline
\hline 
state 1&
\textbf{0.69} {[}0.62; 0.76{]}&
\textbf{0.01} {[}0; 0.03{]}&
\textbf{0.3} {[}0.23; 0.38{]}\tabularnewline
\hline 
state 2&
\textbf{0.02} {[}0; 0.06{]}&
\textbf{0.98} {[}0.94; 0.99{]}&
\textbf{0} {[}0; 0.03{]}\tabularnewline
\hline 
state 3&
\textbf{1} {[}0.91; 1{]}&
\textbf{0} {[}0; 0.08{]}&
\textbf{0} {[}0; 0.04{]}\tabularnewline
\hline
\end{tabular}\end{center}

\subsubsection{Why are the third spikes of unit 4 triplets wrongly attributed to
unit 6?\label{sub:Why-misclassifications}}

\begin{figure}
\begin{center}\includegraphics[scale=0.6,angle=-90]{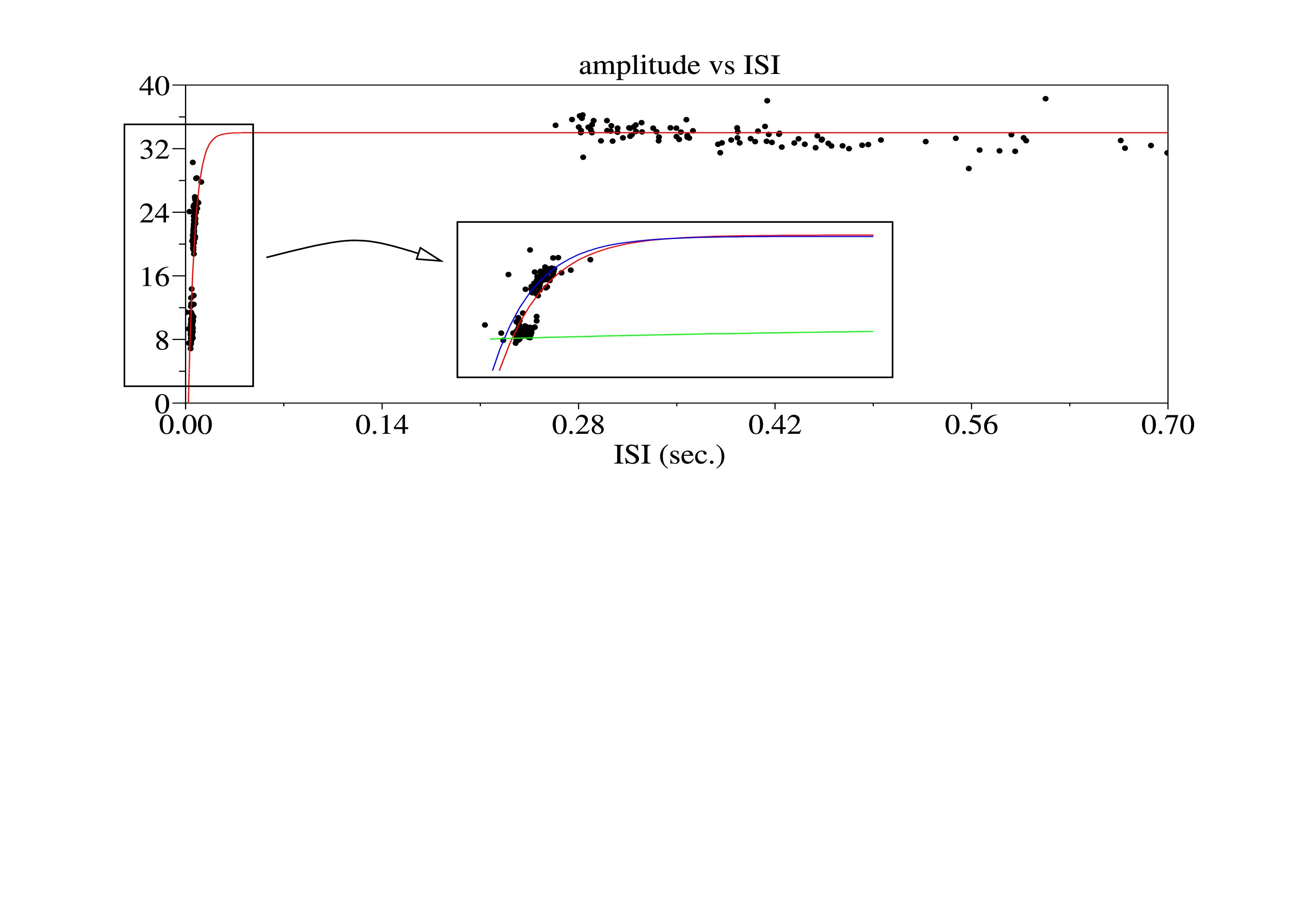}
\end{center}

\caption{\label{fig:Amp-Dyn-Triplet}Amplitude dynamics of spikes within the
triplets of unit~4.
The third spikes of unit~4 triplets (labeled as unit~6 by our algorithm)
and the spikes labeled as unit~4 are pooled together to form the
real spike train of unit~4. The peak amplitudes (after noise whitening)
of these spikes are plotted with respect to the ISI preceding the
spike (black dots). The fitted exponential relaxation is superimposed
(red solid line). Ordinates in units of noise SD. The inset shows
an expanded version of the left hand part of the plot and shows that
the model does not perfectly account for the waveform dynamics of
this unit. This leads to a relatively low likelihood of this spike
train, whereas removing the third spikes (lowest group of points)
of each triplet from unit~4 and attributing them to unit~6 allows
a better global fit (blue solid line for unit~4 and green solid line
for unit~6). The likelihood of the whole is lower in the latter case,
\emph{i.e.} when the third spikes of unit~4 triplets are labeled
as unit~6. This explains why the algorithm fails to fully reconstruct
unit~4 triplets.}
\end{figure}

First of all, the event amplitudes of unit~6 are very similar to
the amplitudes of the third spikes of unit~4 triplets, which considerably
complicates the separation between these two units. In fact, the likelihood
of the data is significantly smaller when these spikes are rightly
labeled as unit~4 than when they are labeled as unit~6, which explains
the output of the algorithm. This is due to the fact that our model
of waveform dynamics is not sufficiently supported by data from unit~4,
so that, with this model, its third spikes in bursts are more likely
to come from unit~6, whose events are of similar amplitude, as shown
in Fig.~\ref{fig:Sorting1}A, site~2. This point is described in
detail in Fig. \ref{fig:Amp-Dyn-Triplet}. Second, as illustrated
on the raw data of site~2 (Fig.~\ref{fig:Sorting1}C1), the third
spikes of these bursts have an overall different waveform (note that
the valley preceding the peak almost disappears). In this case, we
are not dealing with a simple homothetic scaling of the waveform.
That is why we also ran the algorithm using 3 points per site and
per event, instead of the peak amplitude only. This was not sufficient
to correctly label these spikes as unit~4. In fact, using 3 points
per site and per event instead of the peak amplitude did not change
the output of the algorithm in this case. Third, the spikes at stake
here are really small spikes that might even not be detected in other
circumstances. Whatever the spike-sorting method, small events are
always less reliably labeled and the experimentalist has to leave
them out and keep the unambiguous ones. We certainly do not claim
that our method can overcome this limit. In this case, any reasonable
experimentalist who has been dealing with spike-sorting would not
take into account these events.

\end{document}